\documentstyle[12pt]{article}
\title{IMPLICATIONS OF HEAVY TOP}
\author{M.Vysotsky\\
ITEP, Moscow, Russia\\
Lecture at XXIII ITEP Winter School of Physics,\\
Polyani, Zvenigorod, 1995}
\date{}
\begin{document}
\maketitle

\section{Introduction}

Now after discovery of top quark at FNAL collider \cite{1} the third family
of fermions is completed, and the only yet nondiscovered particle of the
Minimal Standard Model (MSM) is Higgs boson. Top quark appeared to be very
heavy, more than 30 times heavier than its partner in electroweak doublet,
$b$-quark:

\begin{equation}
m_t = 176 \pm 8 \pm 10 {\rm GeV} \;\;\;\; CDF\;,
\label{1}
\end{equation}
\begin{equation}
m_t = 199^{+19}_{-21} \pm 22 {\rm GeV} \;\;\;\; DO\;.
\label{2}
\end{equation}
Average:
$$m_t = 180 \pm 12 {\rm GeV}.$$
So it appeared to be  the
heaviest elementary particle known at present. The central
question about the top is: why is it so heavy compared with
the other quarks and
leptons?
What information is hidden under this very specific pattern of quark and
lepton masses? In this lecture we will not discuss this question at
all; our approach will be more practical.  Spontaneous breaking of
gauge invariance in the
electroweak theory leads to a very unusual
manifestation of
the heavy particles: instead of decoupling (phenomena of
power suppression of heavy particles contribution into low energy
observables) contributions of virtual heavy particles are enhanced.
This is the reason why the fact that the top is unusually heavy was
known long before its discovery at FNAL. Large $B-\bar{B}$-mixing
discovered in the eighties \cite{2} signals that $t$-quark is heavy.
Precise measurements of $Z$-boson mass and decay parameters and
$W$-boson mass lead to determination of top mass with
the accuracy close
to that of direct measurements. In this lecture I will mainly deal
with this virtual top effects. In part 2 top implication in $K^0$
system will be discussed; in Part 3 we will go to $B^0$ system. In
Part 4 top implication in $Z$- and $W$-physics will be briefly
considered and, finally, in Part 5 production  and decay of $t$-quark
will be discussed.

\section{$K^0$-mesons}
Electrically neutral pseudoscalar mesons mix with their antiparticles at the
second order of weak interactions. Under this mixing heavy and light
eigenstates are formed (see recent review \cite{3}):
\begin{eqnarray}
P_H = p \mid P^0 > + q\mid \bar{P}^0 >\;,\\
\nonumber
P_L = p\mid P^0 > - q\mid \bar{P}^0 >\;.
\label{3}
\end{eqnarray}
Masses, decay width and coefficients $p$ and $q$ are determined by
the mixing matrix:
\begin{eqnarray}
\left(
\begin{array}{ll}
M - \frac{i}{2}\Gamma  & M_{12} - \frac{i}{2} \Gamma_{12}\\
M^*_{12} - \frac{i}{2} \Gamma^*_{12} & M - \frac{i}{2}
\Gamma
\end{array}
\right)
\left(
\begin{array}{c}
 p \\ q
\end{array}
\right)
= \lambda
\left(
\begin{array}{c}
p\\ q
\end{array}
\right) \;.
\label{4}
\end{eqnarray}
Diagonal matrix elements are equal due to CPT; the violation of CP is
due to the difference of nondiagonal matrix elements. From
(\ref{4}) we get:
\begin{equation}
\lambda_{1,2} = M - \frac{i}{2}\Gamma \pm \sqrt{(M_{12} - \frac{i}{2}
\Gamma_{12})(M^*_{12} - \frac{i}{2} \Gamma^*_{12})}
\label{5}
\end{equation}
\begin{equation}
\Delta M  - \frac{i}{2} \Delta\Gamma = 2 \sqrt{M_{12} - \frac{i}{2}
\Gamma_{12})(M^*_{12} - \frac{i}{2} \Gamma^*_{12})}
\label{6}
\end{equation}
\begin{equation}
\frac{q}{p} = \sqrt{\frac{M^*_{12} - \frac{i}{2}
\Gamma^*_{12}}{M_{12} - \frac{i}{2}\Gamma_{12}}} = 2 \frac{M^*_{12} -
\frac{i}{2} \Gamma^*_{12}}{\Delta M - \frac{i}{2}\Delta \Gamma}
\equiv \frac{1 - \varepsilon}{1 + \varepsilon}
\label{7}
\end{equation}
The phase of the ratio $p/q$ is changed under phase rotation of $P^0$
field: $P^0 \to e^{i \alpha} P^0$, $q/p \to e^{2 i \alpha} q/p$.
So the phase of $q/p$ could be made equal to zero and CP violation
is proportional to $(q/p) - 1 \simeq -2 Re \varepsilon$. Now we have
all general formulas and they should be specified for $K^0 -
\bar{K}^0$ and $B^0 - \bar{B}^0$ systems.

In MSM $K^0 - \bar{K}^0$  transition~~ proceed~~ through the box
diagram shown \\ in Fig. 1.

\vspace{5cm}

\begin{center}
Fig. 1.\\
Feynman diagram responsible for $K^0 - \bar{K}^0$ transition.
\end{center}

When this diagram is calculated in renormalizable $R_{\xi}$ gauge
exchanges of charged components of Higgs doublet should be taken into
account. These are the exchanges that produce  the leading
contribution $\sim m^2_t$. Yukawa coupling of Higgs boson to the
quark bracket $\bar{t}_R(d,s)_L$ is proportional to $m_t$. Taking for
top propagator $G_t = \hat{p}/(p^2 - m^2_t)$ we obtain for the box
diagram under study the following expression:
\begin{equation}
M \sim m^4_t \int \frac{d^4 p p^2}{(p^2 + m^2_t)^2(p^2 + M^2_W\xi)^2}
\sim m^2_t\;,
\label{8}
\end{equation}
which demonstrates non-decoupling behaviour.
The explicit expression for
the box diagram for $m_t \sim m_W$
was for the first time derived in
\cite{4}. Saturating matrix element $<K^0 \mid (\bar{d} s)^2 \mid K^0
>$ by vacuum insertion we get \cite{4}:
\begin{eqnarray}
\label{9}
M_{\bar{K}K} = -\frac{G^2_F}{6\pi^2} f^2_K m^2_K \{\eta_{cc} m^2_c
V^{*2}_{cs} V^2_{cd} + 2m^2_c \ln \frac{m^2_t}{m^2_c} \times
\\
\nonumber
\times V^*_{cs} V_{cd} V^*_{ts} V_{td} \eta_{tc} + m^2_t
I(\frac{m^2_t}{m^2_W}) V^{*2}_{ts} V^2_{td} \eta_{tt}\}\;.
\end{eqnarray}
Here $f_k = 160$ MeV is a constant of $K \to \mu\nu$ decay, $G_F
\simeq 10^{-5}/m^2_p$ is Fermi coupling constant, $m_K$ is $K^0$
mass, factors $\eta_{ij} \approx 0.6$  take into account gluon
exchanges, $V_{ij}$ are the matrix elements of Kobayashi-Maskawa quark
mixing matrix. The first term corresponds to a diagram with two
$c$-quark exchange, the second -- to the case when one quark is
charm, another -- top and, finally, the third term comes from the two
top quarks exchange.  Factor  $I(x)$ calculated in [4] is called in
the literature Inami-Lim factor:
\begin{equation}
I(x) = \frac{4 -
11x + x^2}{4(1-x)^2} - \frac{3x^2 \ln x}{2(1-x)^3}\;,\;\; I(0) =
1\;,\;\;I(4) = 0.6\;,\;\;I(\infty) = 0.25
\label{10}
\end{equation}
{}From the experiments with $K$-mesons mass difference $\Delta m_{LS}$
is known as well as the parameter of CP violation $\varepsilon$.  To
look for top quark contribution we should know the numerical values
of parameters $V_{ij}$.  We will use the following parametrization of
Kobayashi-Maskawa matrix:
\begin{eqnarray}
\left(
\begin{array}{ccc}
V_{ud}  & V_{us}  & V_{ub} \\
V_{cd}  & V_{cs}  & V_{cb} \\
V_{td}  & V_{ts} & V_{tb}
\end{array}
\right)
=
\left(
\begin{array}{ccc}
c_{12} c_{13}   & s_{12} c_{13}  & s_{13}e^{-i\delta}\\
-s_{12} c_{23} - c_{12} s_{23} s_{13} e^{i\delta} &  c_{12} c_{23} -
s_{12} s_{23} s_{13} e^{i\delta} & s_{23} c_{13} \\
s_{12} s_{23} - c_{12} c_{23} s_{13} e^{i\delta}  & -c_{12}s_{23} -
s_{12} c_{23} s_{13} e^{i\delta}  & c_{23}c_{13}
\end{array}
\right) .
\label{11}
\end{eqnarray}

Diagonal elements of CKM matrix equal unity with high
accuracy. Study of strange, charm and beauty particles decay
leads to the following values of CKM matrix elements \cite{5}:
\begin{eqnarray}
\label{12}
\nonumber
\mid V_{us} \mid &=& 0.221 (2) \\
\nonumber
\mid V_{cd} \mid &=& 0.204 (17) \\
\mid V_{cb} \mid &=& 0.040 (5) \\
\nonumber
\mid V_{ub}\mid  &=& 0.003 (1) \nonumber\;.
\end{eqnarray}

Unitarity of CKM matrix makes it possible to determine the values of
$V_{ts}$ and $V_{td}$ which enter (\ref{9}):
\begin{equation}
V^*_{ts} V_{tb} + V^*_{cs} V_{cb} + V^*_{us} V_{ub} = 0\;.
\label{13}
\end{equation}
As $V_{tb}$ and $V_{cs}$ are equal to unity with high accuracy while
$V_{cb} \gg V_{us}^* V_{ub}$, we get:
\begin{equation}
V^*_{ts} \simeq -V_{cb}^*\;.
\label{14}
\end{equation}
Analogously for $V_{td}$ we get:
\begin{equation}
V^*_{td} V_{tb} + V^*_{cd} V_{cb} + V^*_{ud} V_{ub} = 0\;,
\label{15}
\end{equation}
\begin{equation}
V_{td} = -V_{cd} V^*_{cb} - V^*_{ub}
\label{16}
\end{equation}
We are ready now to determine relative importance of top contribution
to $\Delta m_{LS}$. The first term in brackets in  eq.
(\ref{9}) dominate
over the second term. Let us compare the third and the first terms:
\begin{equation}
\frac{tt}{cc} = (\frac{m_t}{m_c})^2 I(4) \frac{V^{*2}_{ts}
V^2_{td}}{V^{*2}_{cs} V^2_{cd}} \leq 3.5 \cdot 10^{-7}
(\frac{m_t}{m_c})^2
\label{17}
\end{equation}
We demonstrate that $t$-quark contribution to the difference of
masses of $K_L$ and $K_S$ mesons is less than 1\%.

Let us note that a charm quark is also not enough to reproduce
the experimental number $(\Delta m_{LS})_{exp} = 3.51(2)
\cdot 10^{-12}$ MeV:
\begin{equation}
(\Delta m_{LS})_{theor} = \frac{\mid M_{K\bar{K}} \mid}{m_K} =
\frac{G^2_F}{6\pi^2} f^2_K m_K \eta_{cc} m^2_c
\mid V_{cs}V_{cd}\mid^2 = 1.1\cdot 10^{-12} {\rm MeV}\;,
\label{18}
\end{equation}
where we substituted $f_K = 160$ MeV, $\eta_{cc} = 0.6\;, \; m_c =
1.3$ GeV, $V_{cs} = 1\;, \; V_{cd} = 0.2$. In
the first equality we use
eq.  (\ref{6}) taking into account smallness of CP-violation in
$K$-mesons $(M_{12} \simeq M_{12}^*\;, \; \Gamma_{12} \simeq
\Gamma^*_{12})$ and the fact that $\Delta M \equiv m^2_H - m^2_L$
since the squares of scalar particle masses enter the Lagrangian.

Now we turn to CP-violation in $K$-mesons. The parameter of
 CP-violation caused by $K^0 - \bar{K}^0$ mixing
$\varepsilon$ is  given by eq.
(\ref{7}).

A nonzero value of $\varepsilon$ induces celebrated $K^0_L$ decay into
two pions. Taking into account that $Im  M_{12} \ll Re M_{12}$,  $Im
\Gamma_{12} \ll Re \Gamma_{12}$ and $Im M_{12} \gg Im \Gamma_{12}$,
we get:
\begin{equation}
\varepsilon = \frac{i Im M_{12}}{\Delta M -
\frac{i}{2} \Delta\Gamma}\;, \;\; \mid \varepsilon \mid = \frac{Im
M_{12}} {2 \sqrt{2} m_K \Delta m_{LS}} = 2.26(2) \cdot 10^{-3}\;,
\label{19}
\end{equation}
where the last number is experimental. Approximate equality $\Delta
m_{LS} \approx \Gamma_S/2$ was used to calculate $\mid \varepsilon
\mid$. From (\ref{9}) and (\ref{11}) we get:
\begin{equation}
Im M_{12} = \frac{G^2_F f^2_K m^2_K \eta}{3\pi^2} s_{12} s_{13}
s_{23} \sin \delta \{m^2_c[\ln(\frac{m^2_t}{m^2_c}) - 1] + I
m^2_t s^2_{23}\}\;.
\label{20}
\end{equation}
It is evident from the last expression that  $t$-quark
contribution  determines the value of $\varepsilon$. The estimate of
the value of CP-violating phase $\delta$ we get substituting in
(\ref{20}) $m_t = 180$ GeV: $\sin \delta \approx 0.7$.

Direct CP-violation in $K^0$ decays takes place in a standard model
due to CKM phase entering $\bar{s}dg$,  $\bar{s}dZ$ and
$\bar{s}d\gamma$ vertices which appear in one loop with $W$-boson
exchange.  Calculation of parameter $\varepsilon'$ which describes
direct CP violation involves $t$-quark loop as well and is very
sensitive to the $m_t$ value.
However the value of $\varepsilon'$  is
sensitive to the values of hadronic  matrix elements and theoretical
prediction has  poor precision.  For $m_t \sim 100 \div 200$ GeV
$\varepsilon'/\varepsilon \sim 10^{-3} \div 10^{-4}$ is predicted.
Experimental situation is controversial as well \cite{6},\cite{7}:
\begin{eqnarray}
NA31:\;\;
Re(\varepsilon'/\varepsilon) = (23.0 \pm 6.5) \times 10^{-4}\;,\\
\nonumber
E731:\;\; Re(\varepsilon'/\varepsilon) = (7.4 \pm 6.0) \times 10^{-4}
\;.
\label{21}
\end{eqnarray}
Future experiments should clarify this discrepancy but cannot
provide test of
MSM.

\section{$B^0$-mesons}
As it was already stated in Introduction large $B^0 -\bar{B}^0$
mixing was the first place where heavy top shows up. The first
evidence for substatial $B^0 - \bar{B}^0$ mixing came from UA1
experiment where overproduction of the same sign dileptons was
observed. In semileptonic decay of $B^0_d(\bar{b}d)$ or
$B^0_s(\bar{b}s)$ meson $l^+$ is produced, while in the decay of
$\bar{B}^0_d$ or $\bar{B}^0_s\;\; l^-$ should appear. Pairs
$b\bar{b}$ are always produced in collisions, so the opposite sign
dileptons should appear when $b\bar{b}$ decays. However
experimentalists saw the same sign dileptons as well. Conclusive
evidence of large $B^0 - \bar{B}^0$ mixing comes form ARGUS
experiment where the decays of $(B_d -\bar{B}_d)$ pairs produced in
the decays of $\Upsilon (4s)$ resonance were studied.

The following relation takes place for the pairs of leptons produced
in $\Upsilon \to B_d\bar{B}_d \to ll X$ decays:
\begin{equation}
r = \frac{N(e^+e^-) + N(e^-e^-)}{N(e^+e^-)} = \frac{x^2}{2 +
x^2}\;\;, \;x = \frac{\Delta M_{B\bar{B}}}{\Gamma_B}\;.
\label{22}
\end{equation}

Measurement of the ratio $r$ determines the difference of masses of
light and heavy mesons. In $\Upsilon(4s)$ decay $B$-mesons are
produced practically at rest, and observation of the space picture of
the $B-\bar{B}$ oscillation is impossible. This space picture of
oscillations was observed later at LEP in the study of $Z \to
B\bar{B}$ decays. The present experimental numbers are \cite{5}:
\begin{eqnarray}
\label{23}
x^{exp}_d = 0.71\pm 0.06 \;,\\
\nonumber
\Delta m^{exp}_{B\bar{B}} = m_{B_H} - m_{B_L} = 3.4(4) \cdot 10^{-13}
{\rm Gev} \;.
\end{eqnarray}
The theoretical expression for this difference can be easily obtained
from (\ref{9}):
\begin{equation}
\Delta m_{B\bar{B}} = \frac{\mid M_{B\bar{B}}\mid}{m_B} =
\frac{G^2_F}{6\pi^2} f^2_B m^2_t m_B I
(\frac{m^2_t}{m^2_W}) \mid V_{td} \mid^2\;.
\label{24}
\end{equation}
Substituting $m_B = 5.3$ GeV, $m_t = 180$ GeV and $f_B = 130$ MeV
\cite{9}, and comparing (\ref{23}) and (\ref{24}) we get:
\begin{equation}
\mid V_{td} \mid \simeq 0.0095\;.
\label{25}
\end{equation}

{}From the unitarity relation (\ref{13}) and the numerical values of
the mixing matrix elements (\ref{12}) we get:
\begin{equation}
0.008 - 0.003 < \mid V_{td} \mid < 0.008 + 0.003
\label{26}
\end{equation}
and we see that the value (\ref{25}) which follows from $B-\bar{B}$
mixing is in good correspondence with the large value of $\sin
\delta$ which we obtain from the study of $CP$-violation in $K$
decays. However, a careful study of the validity of Kobayashi-Maskawa
model of CP-violation is not possible with the present day
experimental accuracy in the measurement of the mixing matrix
parameters.  Considerable progress in this direction should occur
when the study of CP-violation in $B$ mesons will be performed. A
special lecture at this school was devoted to the subject of CP
violation in $B$ mesons  \cite{9}
and we will omit this subject here.  $B_s -
\bar{B}_s$ oscillations were observed at LEP as well.  Theoretically
we have:
\begin{equation}
\Delta m_{Bs} = \frac{G^2_F}{6\pi^2}
f^2_{B_s} m^2_t m_B I(m^2_t/m^2_W) \mid V_{ts} \mid^2 = 0.6 \cdot
10^{-11} {\rm GeV}\;,
\label{27}
\end{equation}
where we assume
$f_{Bs} = f_{Bd} = 130$ MeV. Taking $\tau_{Bs} = 1.34 \cdot 10^{-12}
sec$, we obtain:
\begin{equation}
x_s \simeq 10
\label{28}
\end{equation}

We see that $B_s - \bar{B}_s$ mixing should be very large. This is
the reason why  to measure the theoretically interesting quantity
$x_s$ space picture of $B_s - \bar{B}_s$ oscillatioins should be
studied.
The present experimental bound is \cite{5}:
\begin{equation}
x_{B_s} > 2.0\;.
\label{29}
\end{equation}

Recently CLEO collaboration has measured an inclusive branching
ratio of the $B$-meson decay into a photon and the charmless hadronic
system containing $K$ meson \cite{10} :
\begin{equation}
Br(b \to s\gamma) = (2.32 \pm 0.51 \pm 0.29 \pm 0.32) \cdot 10^{-4}
\;.
\label{30}
\end{equation}
This decay is described by $b \to (u,c,t) \to s$ weak vertex with
photon radiated from the quark or the $W$-boson line.
The expression for this amplitude was found in \cite{11}:
\begin{eqnarray}
\label{31}
M = \frac{G_F}{2\sqrt{2}} \frac{e}{2\pi^2} \sum_i V_{ib}
V^*_{is} F_2^i(x) q_{\mu} \varepsilon_{\nu} \bar{s} \sigma_{\mu\nu}
\frac{1 - \gamma_5}{2} b m_b \;,
\\
\nonumber
F^i_2(x) = Q_i\{[-\frac{1}{4} \frac{1}{x-1} + \frac{3}{4}
\frac{1}{(x-1)^2} + \frac{3}{2} \frac{1}{(x-1)^3}] x -
\\
\nonumber
-\frac{3}{2} \frac{x^2}{(x-1)^4} \ln x \} - x[\frac{1}{2(x-1)} +
\frac{9}{4(x-1)^2} + \frac{3}{2(x-1)^3}] +
\\
\nonumber
+ \frac{3x^3}{2(x-1)^4} \ln x\;, \; Q_i = \frac{2}{3}\;, \; x =
\frac{m^2_i}{m^2_W}\;,
\end{eqnarray}
where $q_{\mu}$ and $\varepsilon_{\nu}$ are the photon momentum and
polarization vector, $\sigma_{\mu\nu}$ is the usual combination of
Dirac matrices, $F(0) = 0,\; F(4) = -0.37,\; F(\infty) = -2/3$. From
(\ref{31}) we obtain:
\begin{equation}
\Gamma_b \to s\gamma = \frac{\alpha G^2_F}{32\pi^4} m^5_b \mid V_{ts}
V^*_{tb} F^t_2 \mid ^2\;.
\label{32}
\end{equation}
With the help of formula for the $b \to c e \nu$ decay probability:
\begin{equation}
\Gamma_b \to c e \nu = \frac{G^2_F m^5_b}{192 \pi^3} [1 - 8
\frac{m^2_c}{m^2_b}] \mid V_{cb} \mid ^2
\label{33}
\end{equation}
we obtain:
\begin{equation}
Br(b \to s\gamma) = \frac{\Gamma(b \to s\gamma)}{\Gamma(b \to c e
\nu)} Br (b \to c e \nu) = \frac{6}{\pi} \alpha
\frac{\mid F^t_2 \mid^2}{1 -
8(\frac{m_c}{m_b})^2} \cdot 10\% \simeq 4.6 \cdot 10^{-4}\;,
\label{34}
\end{equation}
which qualitatively agree with experimental number. Gluon corrections
to the amplitude (\ref{31}) were considered in a number of papers.
They appeared to be large and poorly known.

\section{Top and precision electroweak measurements}
Measurements of the $W$- and the $Z$-boson masses and the $Z$-boson
decay parameters were made with the unprecedental  for the high
energy physics accuracy at CERN, FNAL and SLAC. While in the $K$- and
the $B$-mesons physics the experimental accuracy varies between 1\%
and 10\%
and the theoretical accuracy is of the order of 1\%, in the
$W$ and the $Z$ physics the experimental accuracy is close to 0.2\%,
while the theoretical one varies between 0.1\% and 0.01\% \cite{12}.
Before such accuracy was typical for QED and it is not clear if it
can be reached in the future investigations in HEP. This level
of experimental and theoretical accuracy makes it possible to measure
 radiative corrections to the corresponding amplitudes and especially
 enhanced terms in it. As it is well known there are enhanced
 terms proportional to $m^2_t$ and comparing the theoretical
 expressions with the experimental results the following prediction
 for $t$-quark mass was obtained \cite{13}:
 \begin{equation}
 m_t = 179 \pm 9^{+17}_{-19} {\rm GeV}\;,
 \label{35}
 \end{equation}
 where the first uncertainty is experimental while the second is due
 to unknown value of the Higgs boson mass and corresponds to the
 variation of $m_H$ between 60 GeV and 1 TeV.
 Radiative corrections are proportional to $\frac{m^2_t}{m^2_Z} - \ln
 \frac{m^2_H}{m^2_Z}$.

 I will not discuss here in detail how these bounds on $m_{top}
 $ were obtained as this question was widely discussed in
 literature (see for example lectures in proceedings of the previous
 ITEP Winter School \cite{14}). Let me only stress that
 the conservation
 of gauge currents in QFD (quantum flavordynamics) does not lead to
 decoupling of heavy degrees of freedom unlike the case of QED. In
 QED photon polarization operator looks like:
 \begin{equation}
 \Pi^{\gamma}_{\mu\nu}(q^2) = P^{\gamma}[q^2 g_{\mu\nu} - q_{\mu}
 q_{\nu}]\;,
 \label{36}
 \end{equation}
 where the function $P(q^2)$ is regular at $q^2 = 0$. In QFD we have:
 \begin{equation}
 \Pi^{Z,W}_{\mu\nu} (q^2) = P^{Z,W} (q^2)[g_{\mu\nu} q^2 -
q_{\mu}q_{\nu}]\;,
 \label{37}
 \end{equation}
 where the functions $P^{Z,W}(q^2)$ had a pole at $q^2 = 0$ which is
 created by the goldstone modes admixture to the $Z$- and the
 $W$-bosons.

So, $P^{Z,W}(q^2) \sim m^2_t/q^2$ and corrections proportional to
$m^2_t$ come out. This phenomenon can be easily recognized in
t'Hooft-Landau gauge where the propagators of fictious Higgs degrees
of freedom have the pole at $q^2 = 0$.

\section{Production and decay of $t$-quark}
Top quark was discovered at FNAL $p\bar{p}$-collider. Top was
produced in pair with its antiparticle in quark-antiquark
annihilation via the intermediate gluon or in gluon-gluon fusion.
Experimentally the measured production cross-section appeared to be
larger than theoretically predicted. Being produced top  rapidly
decays to $b$-quark and $W$-boson which, in  turn, decays to two
quark jets or $(l \nu)$ pair. Present accuracy in measurement of
top mass is $\pm 12$ GeV (see eq. (\ref{1}), (\ref{2})). Planned
accuracy in the future LHC experiments is $\pm 3$ GeV while at Next
Linear $e^+e^-$- Collider accuracy up to $\pm 1$ GeV can be
achieved. The  final accuracy in the value of $m_t$ extracted from
$Z$-boson decay parameters measured at LEP should reduce to $\pm 5$
GeV. Comparing this extracted $m_t$ value with the result of direct
measurement one would be able to determine the Higgs boson mass with
the accuracy better than 100 GeV. Direct measurement of the Higgs
boson mass would provide  test of the validity of the minimal
standard model.

Our last topic is top quark decay. For the amplitude of $t \to bW$
decay we have:
\begin{equation}
M_W = \frac{g}{\sqrt{2}} \bar{b} \gamma_{\alpha} \frac{1 +
\gamma_5}{2} t W_{\alpha}\;.
\label{38}
\end{equation}
Calculating the decay width with the help of $W$-boson density
matrix \\
$\rho_{\alpha\beta} = < W_{\alpha} W_{\beta} > =
-(g_{\alpha\beta} - \frac{q_{\alpha}q_{\beta}}{m^2_W})$ we obtain:
\begin{equation}
\Gamma = \frac{g^2}{64\pi} \frac{m^3_t}{m^2_W}  (1 - 3
\frac{m^4_W}{m^4_t} + 2 \frac{m^6_W}{m_t^6})\;.
\label{39}
\end{equation}
Singularity of
the decay width at small $m_W$ is fictious. To study the
small $m_W$ limit one should substitute in (\ref{39})  expressions for
top and $W$ masses through Higgs expectation value:
$m_t = h\eta/\sqrt{2}\;, \; m_W = g \eta/2$. Then for leading at $m_W
\to 0$ term we obtain a regular expression:
\begin{equation}
\Gamma \simeq \frac{g^2}{64\pi} (\frac{\sqrt{2}h}{g})^2 m_t =
\frac{h^2}{32\pi} m_t = \frac{1}{16\pi m_t} \frac{h^2 m^2_t}{2}\;.
\label{40}
\end{equation}

One more way to get leading  in the limit $m_t \gg m_W$ expression
(\ref{40}) exists.
Production of the longitudinal component of the $W^+$-boson
dominates in high energy limit. This longitudinal component
is made from $H^+$. Matrix element for $H^+$ radiation is:
\begin{equation}
M_H = h \bar{b}_L t_R H^+\; ;
\label{41}
\end{equation}
taking square we get:
\begin{equation}
\mid M_H \mid^2 = h^2_t \frac{1}{2} Sp\hat{p}_2 \frac{1 -
\gamma_5}{2} \hat{p}_1 \frac{1 + \gamma_5}{2} = h^2_t(p_1 p_2) =
\frac{h^2_t m^2_t}{2}\;.
\label{42}
\end{equation}
Expression (\ref{40}) follows from (\ref{42}) straightforwardly.
Substituting in (\ref{39}) $m_t = 180$ GeV we obtain:
\begin{equation}
\Gamma_t = 1.7 {\rm GeV}\;.
\label{43}
\end{equation}
Lifetime of $t$-quark is considerably  shorter than the characteristic
hadronic time $\sim 1/100$ MeV. It  means that top decays before top
containing hadron forms, so  unlike the cases with $c$- and
$b$-quarks no top containing hadrons will be discovered in future.

\section{Conclusions}
We demonstrate that in $K$-meson physics the value of $m_{top}$
determine parameter $\varepsilon$ but is inessential for
$\Delta m_{LS}$. The heavy top leads to the large $B-\bar{B}$ mixing.
Precise measurements of the intermediate weak boson properties lead
to the accurate prediction of $t$-quark mass which was confirmed by
measuring mass of $t$-quark produced in $p\bar{p}$ collisions.
Measurement of $m_t$ with several GeV accuracy will provide us with
an estimate of Higgs boson mass.

\vspace{3mm}

\begin{center}
{\bf Acknowledgements}
\end{center}

\vspace{2mm}

I am grateful to M.V.Danilov for suggestion to present these lectures
at ITEP Winter School and to RFFR grant 93-02-14431, ISF grant MRW
300 and INTAS grants 93-3316 and 94-2352 for support.

\end{document}